\RequirePackage{fix-cm}
\documentclass[twocolumn]{svjour3}          % twocolumn
\smartqed  % flush right qed marks, e.g. at end of proof
\usepackage{graphicx}
\journalname{Journal of Materials Science}
\begin{document}

\title{\textit{Ab-initio} calculation of effective exchange interactions, spin waves, and Curie
 temperature in L2$_{\mathrm{1}}$- and  L1$_{\mathrm{2}}$-type local moment ferromagnets}

\titlerunning{Exchange interactions and Curie temperature of L2$_1$- and  L1$_2$-type ferromagnets}
       % if too long for running head

\author{I. Galanakis  \and E. \c{S}a\c{s}{\i}o\u{g}lu}

\institute{I. Galanakis \at
              Department of Materials Science, School of Natural
Sciences, University of Patras,  GR-26504 Patra, Greece \\
              \email{galanakis@upatras.gr}           %  \\
           \and
           E.  \c{S}a\c{s}{\i}o\u{g}lu \at
              Peter Gr\"{u}nberg Institut and Institute for
Advanced Simulation, Forschungszentrum J\"{u}lich and JARA, 52425
J\"{u}lich, Germany and Department of Physics, Fatih University,
34500, B\"{u}y\"{u}k\c{c}ekmece, \.{I}stanbul, Turkey \\
\email{e.sasioglu@fz-juelich.de} }

\date{Received: date / Accepted: date}

\maketitle

\begin{abstract}
Employing  first-principles electronic structure calculations in
conjunction with the frozen-magnon method we study the effective
exchange interactions and spin waves in local moment ferromagnets.
As prototypes we have chosen three L2$_1$-type full Heusler alloys
Cu$_2$MnAl, Ni$_2$MnSn and Pd$_2$MnSn, and the L1$_2$-type XPt$_3$
compounds with X= V, Cr and Mn. We have also included  CoPt$_3$
which is a usual ferromagnet. In all compounds due to the large
spatial separation ($\sim 4$ \AA) of the magnetic transition metal
atoms, the 3\textit{d} states belonging to different atoms overlap
weakly and as a consequence the exchange coupling is indirect,
mediated by the \textit{sp} electrons. Calculated effective
exchange parameters are long range and show RKKY-type
oscillations. The spin-wave dispersion curves are in reasonable
agreement with available experimental data. Using the calculated
exchange parameters we have estimated the Curie temperatures
within both the mean-field and the random-phase approximations. In
local moment ferromagents deviations of the estimated Curie
temperature with respect to the available experimental data occur
when the ground-state electronic structure calculations
overestimate the values of the spin magnetic moments as in
VPt$_3$.
\end{abstract}

\section{Introduction}
\label{intro}

\textit{Ab-initio} (first-principles) electronic structure
calculations based on the density functional theory (DFT) have
been widely employed during the last decades to analyze the
behavior of a constantly increasing number of alloys. The rapid
development of computer resources available to scientists allowed
to study systems with increasing complexity. In the meantime
electronic structure calculations have served to predict new
materials with novel properties susceptible of finding
applications in modern nanotechnology which, very often, were
grown using modern experimental techniques like Molecular Beam
Epitaxy or Pulsed Layered Deposition. First-principles
calculations cover a wide range of materials from metals and
semiconductors to biological systems. Also magnetic materials have
been widely studied using such methods and their properties
connected to the spin of the electrons, \textit{e.g.} spin
moments, magnetic anisotropy energy, magnetooptical properties,
etc,  have been accurately represented.

The drawback of first-principles electronic structures
calculations stems from density functional theory which has been
developed for zero temperature. For realistic applications
temperature plays a crucial role since it greatly affects magnetic
properties, \textit{e.g.} the magnetization of a material drops
with temperature and vanishes at the Curie temperature. Thus it
becomes more and more important to use the results of ab-initio
calculations as the basis for developing complex methodologies to
study the temperature effect on the electronic properties of
materials. To study thermodynamics properties of magnetic
materials the obtained zero temperature electronic structure
results is usually mapped onto classical Heisenberg Hamiltonian,
which can be solved by employing the methods of statistical
physics
\cite{Kubler06,Thoene09,Kurtulus05,Meinert11,Meinert11-2,Bose10,Lezaic06,Entel,Sanyal,Bouzerar}.

So far two methods have been widely used to calculate effective
exchange interactions in magnetic materials within the adiabatic
approximation: (i) the frozen-magnon approach, which is a
reciprocal space method \cite{frozen_magnon} and (ii) the
Lichtenstein's real space method \cite{Lichtenstein}. The common
ground of both approaches is that only the collective excitations,
known as magnons, are taken into account. These excitations
dominate the lower part of the excitation spectra and correspond
to spin-waves which run through the materials and which are
associated to the orientation of the spin moments. Except magnons,
in magnetic materials exist the so-called Stoner excitations which
are associated to excitations of single majority-spin electrons to
unoccupied minority-spin states and thus are accompanied by the
flip of the electrons spin \cite{Glazer84}.  In most systems based
on transition-metal atoms the contribution of the Stoner
excitations to the lower part of the excitation is weak due to the
large exchange splitting of these atoms and in some cases like the
half-metallic ferromagnets they are even separated by a large
Stoner gap from magnons \cite{Laref11}; as Stoner gap  is defined
 the energy difference between the highest occupied majority-spin
state and the lowest unoccupied minority-spin state and thus
corresponds to the lowest possible Stoner excitation energy.

\section{Method and motivation}
\label{sec:2}

As we mentioned above the lower part of the excitation spectra is
dominated by the so-called magnons which involve collective
excitations of the spin magnetic moments. In a simplified picture
we can assume that the magnitude of the spin magnetic moments of
the atoms does not change with respect to the zero Kelvin  value
calculated using ab-initio electronic structure methods but as we
raise temperature atomic spin moments change their orientation in
such a way that the azimuthal angle can be described by a
propagating plane wave characterized by a vector $\mathbf{q}$
belonging in the Brillouin zone. This is the so-called spin-wave
or magnon. If the magnitude of \textbf{q} is small we can assume
that the energy for the creation of the spin-wave is given by the
relation $E(\mathbf{q})=D|\mathbf{q}|^2$ and thus depends only on
the magnitude of the wave vector and not its orientation in the
Brillouin zone; this usually occurs around the $\Gamma$ point as
we will show also later on when presenting the energy-dispersion
of the spin waves. The constant $D$ is referred to as the
"spin-wave stiffness constant". Typical values of $D$ for
transition metals are about 300-600 meV$\:$\AA $^2$ \cite{pajda}.

The ground-state electronic structure calculations are carried out
using the augmented spherical waves method (ASW) \cite{asw} within
the atomic--sphere approximation (ASA) \cite{asa} and the
exchange--correlation potential is chosen in the generalized
gradient approximation \cite{gga}. The method of the calculation
of the exchange constants within the frozen-magnon approximation
has been already presented elsewhere \cite{Sasioglu2004}. Here, to
make the paper reasonably self-contained a brief overview is
given. Notice that since the magnetism is almost exclusively
concentrated on the transition metal site in the compounds under
study (with the exception of CoPt$_3$), as confirmed also from our
calculations presented in the next sections, the equations
presented below are for systems with only one magnetic sublattice
for reasons of simplicity.

To calculate the interatomic exchange interactions we use the
frozen-magnon technique \cite{frozen_magnon} and map the results
of the calculation of the total energy of the helical magnetic
configurations
\begin{equation}
\label{spiral} {\bf s}_n=(\cos({\bf qR}_n)\sin{\theta},\,
\sin({\bf qR}_n)\sin{\theta},\, \cos {\theta})
\end{equation}
onto a classical Heisenberg Hamiltonian
\begin{equation}
\label{hamiltonian}  H_{eff}=-  \sum_{i \ne j} J_{ij} {\bf
s}_i{\bf s}_j
\end{equation}
where $J_{ij}$ is an exchange interaction between two Mn(X) atoms
and ${\bf s}_i$ is the unit vector pointing in the direction of
the magnetic moment at site $i$, ${\bf R}_n$ are the lattice
vectors, ${\bf q}$ is the wave vector of the helix and $\theta$
the polar angle giving the deviation of the moments from the $z$
axis. Within the Heisenberg model (\ref{hamiltonian}) the energy
of the frozen-magnon configurations can be represented in the form
\begin{equation}
\label{eq:e_of_q}  E(\theta,{\bf q})=E_0(\theta)-\sin^{2}\theta
J({\bf q})
\end{equation}
where $E_0$ does not depend on {\bf q} and $J({\bf q})$ is the
Fourier transform of the  parameters of exchange  interaction
between pairs of  magnetic atoms:
\begin{equation}
\label{eq:J_q}  J({\bf q})=\sum_{{\bf R} \neq 0} J_{0{\bf
R}}\:\exp(i{\bf q\cdot R}).
\end{equation}
Calculating $ E(\theta,{\bf q})$ for a regular ${\bf q}$-mesh in
the Brillouin zone of the crystal and performing back Fourier
transformation one gets exchange parameters $J_{0{\bf R}}$ between
pairs of the magnetic atoms.

First, the Curie  temperature is estimated within the mean-field
approximation (MFA). For the case of a material with one magnetic
sublattice (for the multi-sublattice case see Refs.
\cite{Sasioglu2004,Anderson}) the equation is
\begin{equation}
\label{eq_system} k_B T_\mathrm{C}^\mathrm{MFA} =
\frac{2}{3}\sum_{{\bf R}\neq 0}J_{0{\bf R}}.
\end{equation}
Within the random phase approximation (RPA) the Curie temperature
is given by the relation \cite{pajda}
\begin{equation}
\label{eq_RPA} \frac{1}{k_\mathrm{B}T_\mathrm{C}^\mathrm{RPA}}=
\frac{6\mu_\mathrm{B}}{M}\frac{1}{N}\sum_q\frac{1}{\omega(\mathbf{q})},
\end{equation}
\noindent where $\omega(\mathbf{q})=4/M[J(0)-J(\mathbf{q})]$ is
the energy of spin-wave excitations, $\mu_{\mathrm{B}}$ is the
Bohr magneton, $N$ is the number of $\mathbf{q}$ points in the
first Brillouin zone, and $M$ is the atomic magnetic moment.
Within the RPA formalism as implemented in our method it is not
possible to take into account multiple magnetic sublattices
contrary to the implementation of MFA. Finally we should note that
MFA corresponds to an equal weighting of the low- and high-energy
spin-wave excitations leading to an overestimation of the
experimental Curie temperature contrary to RPA where the
lower-energy excitations make a larger contribution to the Curie
temperature leading to more realistic values
\cite{pajda,SasiogluH1,SasiogluH2}.

We have applied the frozen-magnon approximation to a variety of
systems. The obtained Curie temperatures were found to be in good
agreement with available experimental data when the Stoner gap was
large enough so that magnons are well-separated from Stoner
excitations in the excitation spectra. Such systems are the
half-metallic ferro-, ferri- and antiferro-magnetic compounds
where the majority-spin band is metallic and the minority-spin
band is semiconducting. The gap at the Fermi level in
minority-spin band leads to a large value of the Stoner gap.
Half-metallic systems where we have applied our method include the
diluted magnetic semiconductors \cite{frozen_magnon},
ferromagnetic Heusler alloys like NiMnSb and Co$_2$MnSi
\cite{SasiogluH1,SasiogluH2,Galanakis08,Sandratskii07},
half-metallic antiferromagnets \cite{Galanakis07,Galanakis11},
transition metal pnictides and chalcogenides
\cite{Sandratskii07,Sasioglu05,Hortamani08,Hortamani09}, and
half-metallic sp-electron ferromagnets
\cite{Laref11,Gao07,Gao07-2}. Moreover these results have served
to study several other thermodynamics properties like the
temperature dependance of the magnetization in half-metallic
antiferromagnets \cite{Sasioglu09} or the thermal properties of
the Ni sublattice in NiMnSb \cite{Sandratskii08}. Local
ferromagnets which have been studied include the Mn-based semi and
full Heusler alloys like NiMnSb and Ni$_2$MnGa
\cite{Sasioglu2004,SasiogluH3,SasiogluH4}.  As shown in Ref.
\cite{Kubler83} the Mn $d$ states are strongly delocalized and
hybridize with the $d$ and $p$ electrons of the neighboring
transition-metal and sp atoms, respectively. The localization of
the spin magnetic moment comes from the fact that minority-spin
electrons are almost excluded from the Mn site while almost all
majority-spin states are occupied.

\begin{table*}
\caption{Lattice parameter $\alpha$ (in \AA ) used in the
calculations, atom-resolved and total spin magnetic moments (in
$\mu_B$), and both calculated and experimental Curie temperatures
(in K) for the local moment ferromagnets under study and CoPt$_3$
which is a usual ferromagnet. Note that for the calculation of the
Curie temperature we have employed both the mean-field (MFA) and
random-phase (RPA) approximations and we have considered only
Mn-Mn or X-X interactions except the cases of MnPt$_3$ and
CoPt$_3$  for which we present in parenthesis also the values with
the MFA taking into account also the Mn(Co)-Pt and Pt-Pt
interactions. Lattice constants and experimental data come from
Refs \cite{Pickart63}, \cite{Menzinger66}, \cite{Kawakami79} ,
\cite{Tajima77}, \cite{Noda76} and \cite{Paul79}.}\label{table1}
\begin{tabular}{lcccccccc}
\hline\noalign{\smallskip}
 Compound & $a$  &
$m_\textmd{(Cu,Ni,Pd)}$ & $m_\textmd{Mn}$ & $m_\textmd{(Al,Sn)}$ &
$m_\textmd{Total}$  & $T_\textmd{c}^{\textmd{MFA}}$ &
$T_\textmd{c}^{\textmd{RPA}}$&
$T_\textmd{c}^{\textmd{Exp}}$ \\
\noalign{\smallskip}\hline\noalign{\smallskip}
Cu$_2$MnAl &  5.95 &  0.02 &  3.67 &  -0.11 &  3.60 &      970 &  635 &  603  \\
Ni$_2$MnSn &  6.05 &  0.21 &  3.73 &  -0.06 &  4.09 &      320 &  254 &  360  \\
Pd$_2$MnSn &   6.38 &  0.07 &  4.08 &  -0.06 &  4.16 &     252 &
178 &  189
 \\ \hline \hline
 Compound &  $a$  & &
$m_\textmd{X}$ & $m_\textmd{Pt}$ & $m_\textmd{Total}$ &
$T_\textmd{c}^{\textmd{MFA}}$ & $T_\textmd{c}^{\textmd{RPA}}$&
$T_\textmd{c}^{\textmd{Exp}}$ \\
\noalign{\smallskip}\hline\noalign{\smallskip}
VPt$_3$   & 3.870 & & 1.564&   -0.089&  1.296 &  740& 631& 290\\
CrPt$_3$  &3.877& &2.871   &-0.071 & 2.658 &  834 &662 &494\\
MnPt$_3$ &3.898 &&3.863 &0.131 &4.257 &  406(414) &317 &390\\
CoPt$_3$   &3.854& &1.776 &   0.268 &2.580   &145(220) &130 &288\\

\noalign{\smallskip}\hline
\end{tabular}
\end{table*}

In this article we consider two classes of local-moment
ferromagnets: (i) the full-Heusler alloys Cu$_2$MnAl, Ni$_2$MnSn
and Pd$_2$MnSn crystallizing in the L2$_1$ structure, and (ii) the
XPt$_3$ alloys in the L1$_2$ structure with X= V, Cr, Mn and Co.
Both lattice structures are cubic and have the same symmetry group
as the fcc lattice. The Mn-based compounds have attracted a lot of
interest due to the localized nature of Mn spin moments
\cite{Kubler83}. The XPt$_3$ compounds on the other hand have also
attracted considerable attention both theoretically
\cite{Shirai95,Kulatov96,Galanakis02,Kwon03,Iwashita96,Galanakis00,Oppeneer02,Galanakis01}
and experimentally
\cite{Pickart63,Menzinger66,Bacon63,Kawakami79,Grange98,Imada99,Lange98,Maret05}
due to the variety of magnetic order which they exhibit and due to
the large magnetocrystalline anisotropy shown by CrPt$_3$
\cite{Oppeneer02} and CoPt$_3$ \cite{Grange98} which makes the
latter alloys attractive for magnetic storage devices. Moreover
these systems are prototypes for studying the induced magnetism at
the Pt site due to the strong hybridization between the
$d$-orbitals of the transition-metal atoms and the platinum atoms
\cite{Galanakis02}. Note that when X=V, Cr, Mn or Co the compounds
are ferromagnets (the compounds with V, Cr and Mn are also local
moment ferromagnets) \cite{Pickart63,Menzinger66,Kawakami79},
while the FePt$_3$ shows antiferromagnetism \cite{Bacon63} and
this is why we have not included it in our study.

Due to the variety of compounds under investigation and to help
the reader we present in the next section first the results for
the three Heusler compounds and MnPt$_3$ for which the magnetism
is concentrated at the Mn site and in Section \ref{sec:4} we
compare the obtained results for MnPt$_3$ with the rest of the
XPt$_3$ alloys.

\section{Mn-based alloys}
\label{sec:3}

Alloys where coexist Mn atoms with sp atoms and/or late transition
metal atoms are susceptible of being local ferromagnets; the
majority-spin electronic band of Mn atoms is almost completely
filled while the charge at the minority-spin band is negligible
and as a result minority-spin electrons are excluded from the Mn
site leading to a localization of the Mn spin moment. Local spin
magnetic moments do not necessarily mean that Mn $d$-electrons are
localized and as shown by K\"ubler \textit{et al.} in Ref.
\cite{Kubler83} the Mn $d$-orbitals are strongly delocalized due
to their strong hybridization with the neighboring $d$ and $p$
orbitals. Prototypes for local ferromagnetism are the full Heusler
alloys like Cu$_2$MnAl or Ni$_2$MnAl which have been extensively
studied in \cite{Kubler83}. Here we will focus on compounds for
which the experimental dispersion curve of the  magnons  for
crystals is known and thus can be compared to our calculated
results. We have chosen three Heusler alloys crystallizing in the
L2$_1$ structure which consists of four interpenetrating fcc
sublattices: Cu$_2$MnAl which has been experimentally studied by
Tajima et al in 1977 \cite{Tajima77}, and Ni$_2$MnSn and
Pd$_2$MnSn studied by Noda and Ishikawa in 1976 \cite{Noda76}. We
have also included in our study the cubic MnPt$_3$ alloy studied
by by Paul and Stirling in 1979 \cite{Paul79}  which crystallizes
in the so-called L1$_2$ structure. For the three Heusler compounds
we have used the experimental lattice constants and for MnPt$_3$
the lattice constant given in Ref. \cite{Shirai95} (see Table
\ref{table1} for the values). For all calculations we have used a
dense 30$\times$30$\times$30 grid in the Brillouin zone to carry
out the needed integrations in the reciprocal space.

For all four compounds the spin magnetic moment is localized at
the Mn site as shown in Table \ref{table1} and the other sites
carry a negligible spin moment with respect to the Mn atoms. The
Mn spin moment varies from 3.67 to 4.08 $\mu_B$ and the small
variation is due to the effect of the local environment (the
degree of hybridization with the orbitals of the neighboring
atoms). We do not present for the Heusler alloys the atom-resolved
and total density of states (DOS) since it is similar to the ones
in Ref. \cite{Kubler83}. Relevant to our discussion is the
Mn-resolved DOS, which varies little between the four compounds
under study and in Fig. \ref{fig3} we present the total and
Mn-resolved DOS in MnPt$_3$. The Mn majority-spin states are
almost completely occupied while the minority states are
completely empty; there is a very small weight of minority-spin
occupied states due to the influence of the neighboring Pt atoms
but this  does not change the overall picture.

\begin{figure*}
\includegraphics[width=17cm]{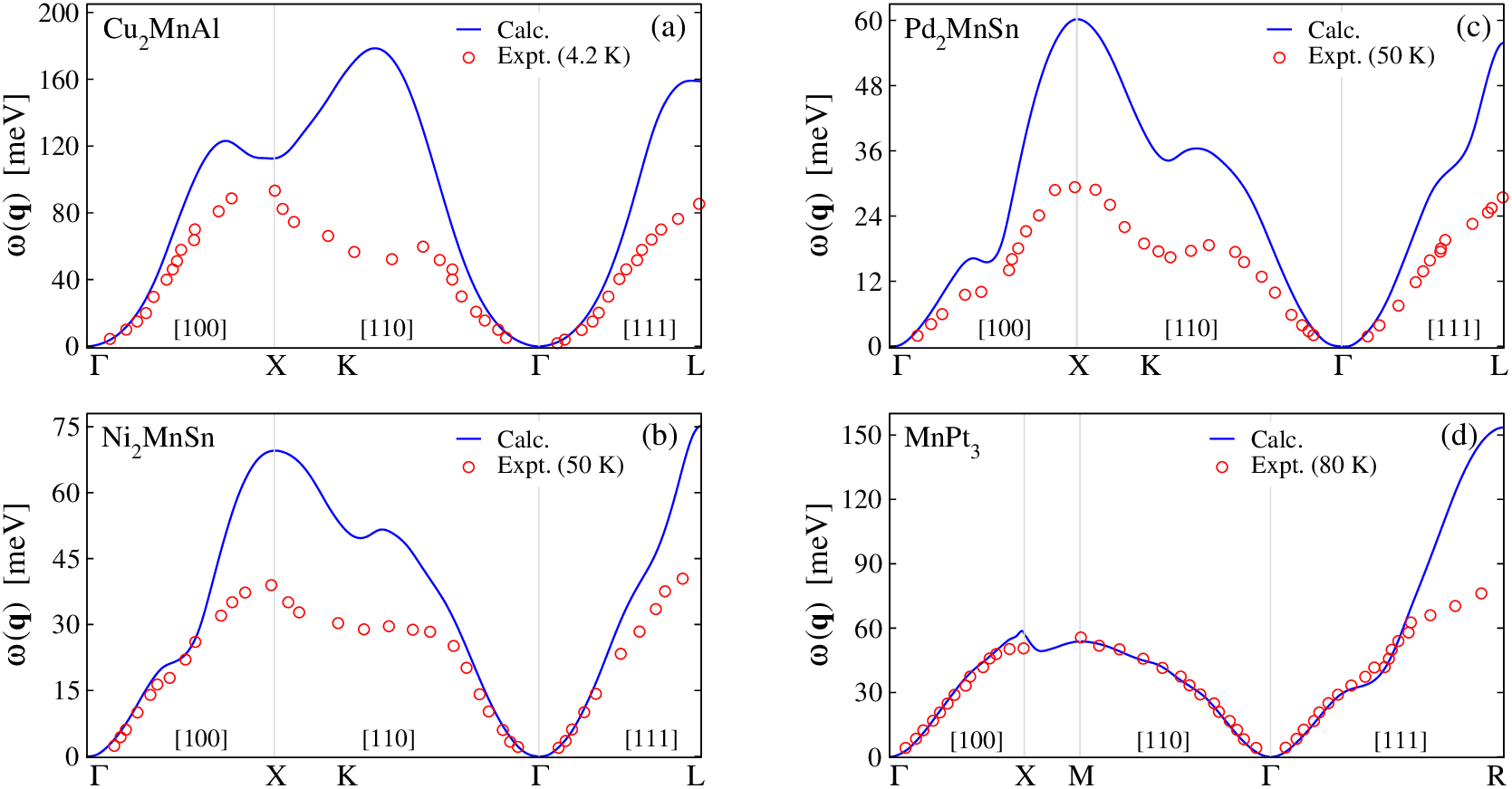}
\caption{Calculated (solid lines) spin-wave dispersion curves in
the first Brillouin zone along high-symmetry axis for the
local-moment ferromagnets. The experimental data (open circles)
have been reproduced from Ref. \cite{Tajima77} for Cu$_2$MnAl,
Ref. \cite{Noda76} for Ni$_2$MnSn and Pd$_2$MnSn and Ref.
\cite{Paul79} for MnPt$_3$. Note that experimental data have been
measured at different temperatures. } \label{fig1}
\end{figure*}

Although a real Stoner gap does not exist as in half-metals, due
to the \textit{sp-d} mixing the majority-spin DOS below the Fermi
level in the L2$_1$-type compounds (see Ref. \cite{SasiogluH4})
and the minority-spin DOS just above the Fermi level in the
L1$_2$-type compounds (see Fig. \ref{fig3}) are almost negligible;
an exception occurs as shown in Fig. \ref{fig3} for CoPt$_3$ which
is not a local moment ferromagnet and substantial part of the Co
minority-spin states are occupied. Therefore, the single-particle
spin-flip Stoner excitations for the local ferromagnets make a
small contribution to the total excitation spectra at low
energies, where collective spin-wave modes dominates. As a result,
well defined spin waves exist throughout the Brillouin zone with
small damping as observed in experiments (see Fig.\ref{fig1}),
which also justifies the use of the frozen-magnon method in the
calculation of the spin-wave dispersion and  the exchange
interactions. Indeed, recent time-dependent DFT calculation of
spin-wave dispersion of Cu$_2$MnAl gave very similar results to
the one obtained from the adiabatic approximation, \textit{i.e.},
Liechtenstein formula \cite{TDDFT}.

\begin{table}
\caption{Calculated ($D_\mathrm{th}$) and experimental
($D_\mathrm{th}$) spin-wave stiffness constant for four local
ferromagnets. Experimental data come from Refs \cite{Tajima77},
\cite{Noda76} and \cite{Paul79}.}\label{table2}
\begin{tabular}{lccc}
\hline\noalign{\smallskip}
 Compound & Structure & $D_{\textrm{th}}$(meV\AA $^2$) & $D_{\textrm{ex}}$ (meV\AA $^2$) \\
\noalign{\smallskip}\hline\noalign{\smallskip}
Cu$_2$MnAl & L2$_1$ &   240 &  167    \\
Ni$_2$MnSn & L2$_1$ &   166 &  154    \\
Pd$_2$MnSn & L2$_1$ &   138 &  123    \\
MnPt$_3$   & L1$_2$ &   260 &  270   \\
 \noalign{\smallskip}\hline
\end{tabular}
\end{table}

In Fig. \ref{fig1} we present the calculated spin-waves dispersion
energy $\omega(\mathbf{q})$ curves and we compare them with the
available experimental data from Refs
\cite{Tajima77,Noda76,Paul79} employing the neutrons diffraction
technique. Note that experiments have been carried out for
different temperature from 4.2 K for Cu$_2$MnAl up to 80 K for
MnPt$_3$. A close look at the results in Ref. \cite{Noda76}
reveals that as the temperature rises for the  same $\mathbf{q}$
experiments give a smaller $\omega(\mathbf{q})$ value. Overall the
calculated dispersion curves follow the behavior of the
experimental data with the minimum at the $\Gamma$ point. For the
three Heusler alloys our calculations give larger values of the
dispersion energies $\omega(\mathbf{q})$, except close to the
$\Gamma$ point where calculations accurately represent the
experimental data. For MnPt$_3$ calculated and experimental data
fall one on top of the other except close to the R point, but this
agreement is misleading since the experiment has been carried out
at 80 K much higher than for Cu$_2$MnAl. The good representation
of the experimental data by the calculations can be traced at the
spin-wave stiffness constants $D$ which are presented in Table
\ref{table2}. As we mentioned above the $D$ is the constant
connecting the energy needed for  the creation of the spin-wave
when the magnitude of the wave-vector $\mathbf{q}$ is small as it
occurs around the $\Gamma$ point. We get the largest discrepancy
for Cu$_2$MnAl where theory and experiment give values of 240 and
167 meV \AA $^2$, respectively, while for the other three
compounds theory overestimates the value of $D$ by less than 15
meV \AA $^2$.

\begin{figure}
\includegraphics[width=\columnwidth]{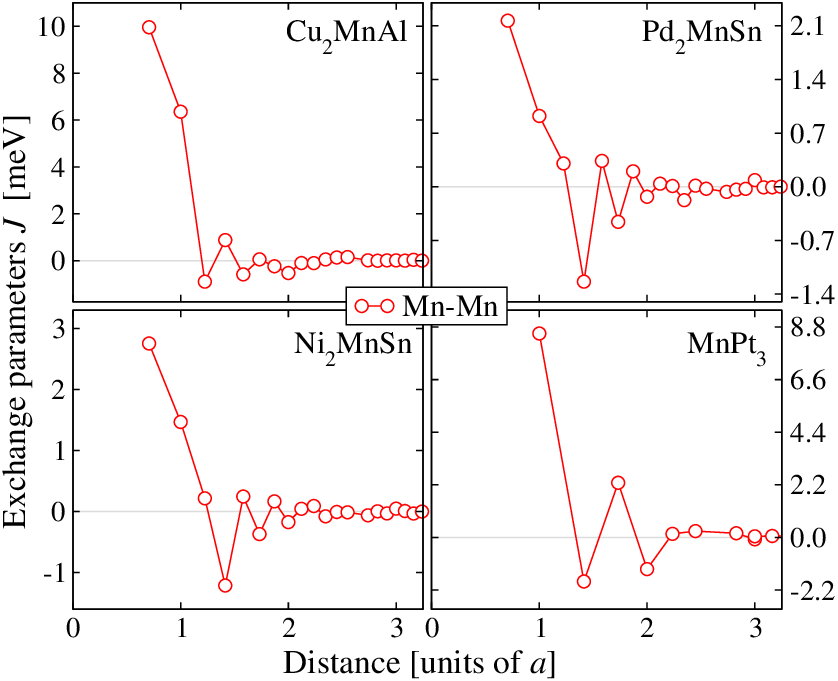}
\caption{Mn-Mn exchange constants in the four Mn-based
local-moment ferromagnets under study as a function of their
distance. The distance is given in units of the lattice parameter:
$\alpha$.} \label{fig2}
\end{figure}

From the spin-waves dispersion energies we can calculate the
exchange constants describing the effective Mn-Mn exchange
interactions as discussed in the previous section. In Fig.
\ref{fig2} we present their values as a function of the distance
between the Mn atoms. In all compounds due to the large spatial
separation ($\sim 4$ \AA) of the magnetic transition metal atoms
the 3\textit{d} states belonging to different atoms  overlap
weakly and as a consequence the exchange coupling is indirect,
mediated by the \textit{sp} electrons. A very nice discussion on
the exchange coupling in local moment ferromagnets cam be found in
Ref. \cite{SasiogluH4}. As seen from Fig. \ref{fig2} the
calculated effective exchange parameters are long range and show
RKKY-type oscillations. All compounds show tendency to
ferromagnetism but the different behavior of the exchange
constants as a function of the distance strongly influences the
Curie temperatures and thus the temperature behavior of the
magnetization. In Table \ref{table1} we have also gathered the
Curie temperature within both the MFA and RPA approaches together
with the experimental values from Refs
\cite{Tajima77,Noda76,Paul79}. In MFA all magnons contribute
equally to the Curie temperature, while in RPA the magnons with
the lower energy have a higher contribution as can be deduced by
the equations in the previous section. Thus RPA is expected to
give a more accurate description of the Curie temperature. For
Cu$_2$MnAl, as shown in Fig. \ref{fig2}, the coupling between Mn
nearest and next-nearest neighbors is strongly ferromagnetic
reaching a value of 10 meV for nearest neighboring Mn atoms and it
almost vanishes for more distant Mn atoms. Thus the ferromagnetism
in Cu$_2$MnAl is extremely stable and the RPA calculated Curie
temperature is 635 K well above the room temperature and close to
the experimental value of 603 K. The Curie temperature calculated
within MFA is more than 50\%\  larger than the RPA value. In both
Pd$_2$MnSn and Ni$_2$MnSn the interactions between a Mn atom and
up to its third neighbors are positive while between the Mn atoms
fourth neighbors it is negative; for further distance it is much
smaller. Note also that the scale in the vertical axis is
different than Cu$_2$MnAl and the maximum value is about 2-3 meV.
Since the number of fourth neighbors is large with respect to the
closest neighbors, ferromagnetism in these two Heusler compounds
is not as robust as in Cu$_2$MnAl and the Curie temperature is
close to the room temperature for Ni$_2$MnSn and below room
temperature for Pd$_2$MnSn for which the antiferromagnetic
interaction between Mn atoms fourth neighbors is larger in
magnitude. MnPt$_3$ exhibits strong ferromagnetic coupling between
Mn atoms nearest neighbors while for Mn second and third neighbors
the coupling is same in magnitude but of opposite sign. Thus the
nearest neighbors interaction stabilizes ferromagnetic order with
an experimental Curie temperature around 390 K. Note that for
Ni$_2$MnSn, Pd$_2$MnSn and MnPt$_3$ where the tendency to
ferromagnetism is not as strong as for Cu$_2$MnAl MFA gives values
for the Curie temperature closer to the experimental ones than PRA
although the latter one is in-principle more exact. This behavior
is contrary to the behavior of the compounds regarding the
spin-wave stiffness constants, since for the latter only the
region around the $\Gamma$ point is important while for the
calculation of the Curie temperature we take into account the
dispersion of the spin-wave energy in the whole Brillouin zone.

\begin{figure}
\includegraphics[width=\columnwidth]{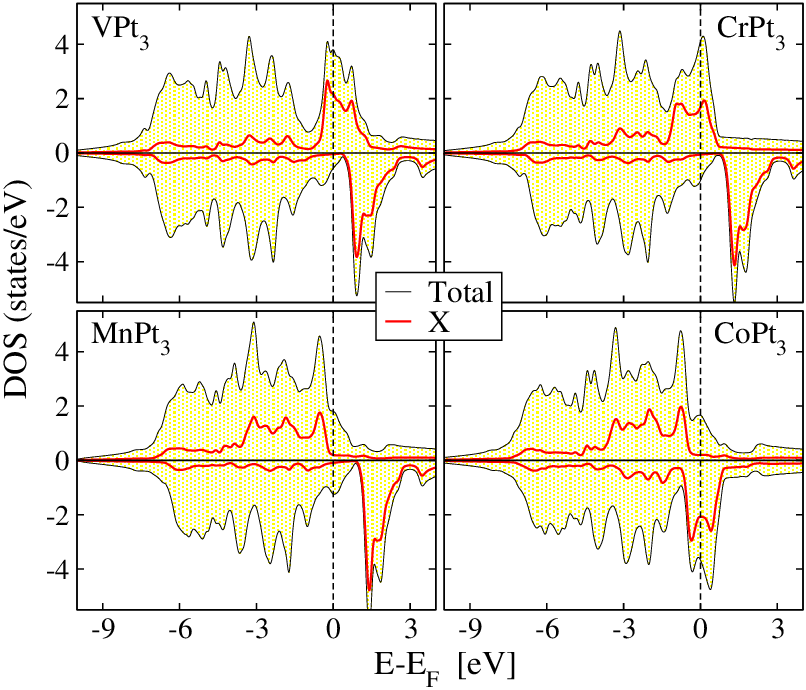}
\caption{Total and transition-metal resolved DOS for the six cubic
XPt$_3$ ferromagnets under study. Positive DOS values are assigned
to the majority spin electrons and negative DOS values to the
minority spin electrons.} \label{fig3}
\end{figure}

\section{XPt$_3$ alloys}
\label{sec:4}

\begin{figure}
\includegraphics[width=\columnwidth]{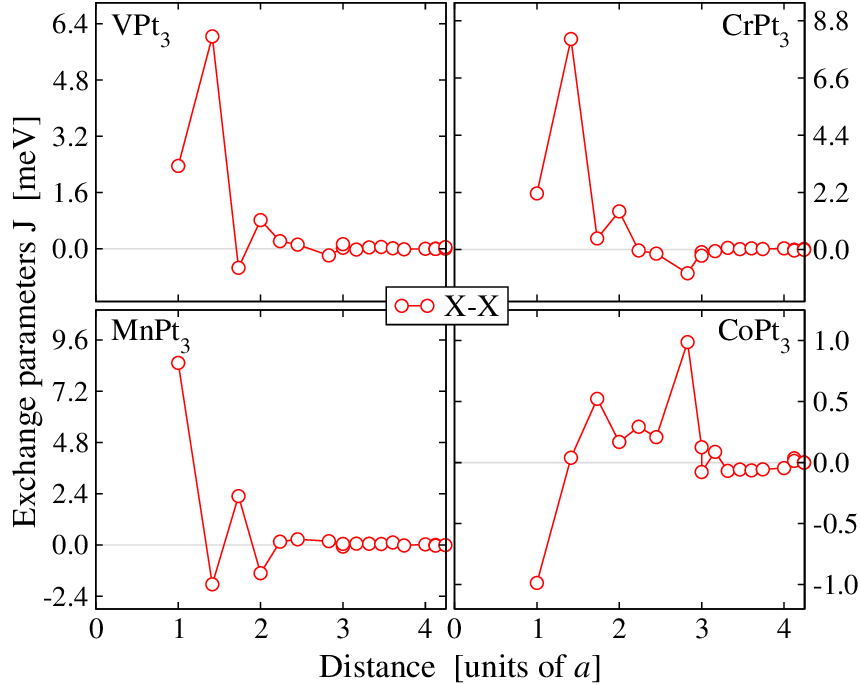}
\caption{X-X exchange constants for the XPt$_3$ alloys under study
as a function of their distance. Details as in Fig. \ref{fig2}.}
\label{fig4}
\end{figure}

In Fig. \ref{fig3} we present the X-resolved and total DOS for all
compounds under study and in Fig. \ref{fig4} the calculated
exchange constants. In the lower part of Table \ref{table1} we
have gathered the lattice constants used for the calculations, the
atom-resolved and total spin magnetic moments, the Curie
temperature within both MFA and RPA as well as the experimental
Curie temperature. We will start our discussion from the DOS of
the local moment ferromagnets (X= V, Cr and Mn) presented in Fig.
\ref{fig3}. As we move from V to Cr and then to Mn the extra
electrons occupy majority-spin states and the minority-spin states
are above the Fermi level. This is also reflected on the
X-resolved spin magnetic moments in Table \ref{table1} where from
V to Cr the transition-metal spin magnetic moment increases by
$\sim$1.3 $\mu_B$ and from Cr to Mn by about 1 $\mu_B$. Since the
majority-spin states in VPt$_3$ and CrPt$_3$ are not completely
occupied these alloys should be classified as weak ferromagnets
contrary to MnPt$_3$ where all Mn majority-spin states are
occupied. In all three alloys the Pt atoms have induced spin
magnetic moments two orders of magnitude smaller than the X atoms
and thus are not relevant for the discussion of the Curie
temperature. To make this point clear we have calculated for
MnPt$_3$ the Curie temperature within MFA taking into account only
the Mn-Mn interactions and all Mn-Mn, Mn-Pt and Pt-Pt interactions
and the value changed from 406 K to 414 K and thus Mn-Mn
interactions are the dominant ones and the Pt sublattice can be
neglected in the latter discussion. The MFA values for both
VPt$_3$ and CrPt$_3$ are much larger than the RPA ones by more
than 100 K as expected form the discussion in Section \ref{sec:2}.
But as we move from Mn to Cr and then to V the discrepancy between
the RPA values and the experimental ones increases. The difference
should be traced in the experimental spin magnetic moments
associated to these Curie temperatures. In Ref. \cite{Kawakami79}
VPt$_3$ shows a Curie temperature of 290 K and Vanadium atoms a
spin magnetic moment of about 1.0 $\mu_B$; both values are much
smaller than our calculated values of 631 K and 1.56 $\mu_B$. For
CrPt$_3$(MnPt$_3$) the experimental values in Ref.
\cite{Pickart63} are 494(390) K and 2.33(3.60) $\mu_B$. Thus as we
move from V to Mn the calculated spin magnetic moments come closer
to the experimental values and so do the Curie temperatures. The
large discrepancy exhibited for V is a well-known problem of
density-functional-theory-based calculations and can be traced
also to other V-based compounds \cite{Galanakis01}. Regarding the
exchange parameters in Fig. \ref{fig4} V-V and Cr-Cr interactions
present a similar picture with nearest and next-nearest neighbors
being ferromagnetically coupled and the coupling is stronger for
second than first neighboring transition metal atoms. For MnPt$_3$
as we discussed in Section \ref{sec:3} nearest neighbors present a
strong ferromagnetic coupling while Mn-Mn next-nearest neighbors
are antiferromagnetically coupled.

In CoPt$_3$,  like MnPt$_3$, all transition-metal majority-spin
states are occupied but contrary to the latter alloy  significant
part of the Co minority-spin states is also occupied and CoPt$_3$
cannot be classified as a local moment ferromagnet. The exchange
splitting between the Co majority- and minority-spin electronic
bands is smaller than for Mn in MnPt$_3$ due to the extra
electrons which fill also minority-spin states. Moreover Pt atoms
carry now a significant induced spin magnetic moment of about 0.27
$\mu_B$ about one sixth of the Co spin magnetic moment and thus
both Co-Pt and Pt-Pt interactions are now important to calculate
the temperature dependent properties. The Co-Co exchange constants
presented in Fig. \ref{fig4} show a peculiar behavior since they
are negative for nearest-neighbors, vanishing for next-nearest
neighbors and positive for further neighbors. This behavior stems
from the strong interplay between the ferromagnetic RKKY and
antiferromagnetic superexchange interactions
\cite{SasiogluH3,SasiogluH4}  and although finally CoPt$_3$ is
ferromagnetic the Curie temperature is smaller than the room
temperature. Experiments in Ref. \cite{Menzinger66} give a value
of 288 K while our RPA and MFA values taking into account only the
Co-Co interactions are 130 K and 145 K respectively. However, the
multi-sublattice MFA gives a value of 220 K considerably larger
than the 145 K calculated taking into account only the Co-Co
interactions. The multisublattice RPA is expected to give a
T$_\mathrm{c}$ below 200 K. The discrepancy for CoPt$_3$ should be
attributed to the underestimation of the exchange interactions in
the frozen-magnon method in the case of small magnetic moments
\cite{Patrick}. It is more likely that the Co-Pt exchange
interactions are underestimated due to the small Pt magnetic
moments.

\section{Conclusions}

Combining first-principles electronic structure calculations and
the frozen-magnon approximation we have studied the thermodynamic
properties of two classes of local moment ferromagnets (i) the
Heusler alloys Cu$_2$MnAl, Ni$_2$MnSn and Pd$_2$MnSn, and (ii) the
cubic XPt$_3$ compounds with X= V, Cr and Mn. The localization of
the spin magnetic moment of the transition-metal atoms arises from
the exclusion of the minority-spin electrons from these sites.  We
have also included in our study CoPt$_3$ which is a usual
ferromagnet.

In all compounds under study due to the large spatial separation
($\sim 4$ \AA) of the magnetic transition metal atoms, the
3\textit{d} states belonging to different atoms overlap weakly and
as a consequence the exchange coupling is indirect,  mediated by
the \textit{sp} electrons. In the case of local ferromagnets a
close look at the density of states (DOS) revealed that due to the
\textit{sp-d} mixing either the majority-spin DOS below the Fermi
level or the minority-spin DOS just above the Fermi level are
almost negligible, and therefore, the single-particle spin-flip
Stoner excitations make a small contribution to the total
excitation spectra at low energies, where collective spin-wave
modes (magnons) dominate and the use of the frozen-magnon
approximation in our calculations is justified. This reasoning is
not applicable to CoPt$_3$ where no localization of the spin
magnetic moment occurs. The calculated effective exchange
parameters are long range and show RKKY-type oscillations. The
Curie temperature is estimated within both the mean-field and the
random-phase approximations and our calculations show that when
the spin magnetic moment is concentrated at the transition-metal
atom only the exchange constants between these atoms determine the
value of the Curie temperature. In general, the calculated Curie
temperatures, exchange constants and spin-wave dispersion curves
are in fair agreement with the available experimental data.

Overall we can conclude that the main criterion for the
application of the frozen-magnon approximation is either the large
value of the Stoner gap as in half-metallic ferromagnets (see
discussion in Section \ref{sec:2}) or a strong exchange splitting
between the transition-metal majority- and minority-spin
electronic bands as for the  local moment ferromagnets under study
in this article. Deviations of the estimated Curie temperature
with respect to the  experimental data occur when the ground-state
electronic structure calculations overestimate the values of the
spin magnetic moments as in VPt$_3$ or when the exchange splitting
is not strong enough as in the usual ferromagnet CoPt$_3$.

\begin{acknowledgements}
Fruitful discussions with L.M. Sandratskii are acknowledged.
\end{acknowledgements}

\end{document}